\documentclass[12pt]{article}
\usepackage{amsmath}
\usepackage{graphicx}
\usepackage{color}
\begin{document}
\baselineskip=18 pt
\begin{center}
{\large{\bf Spin-zero system of DKP equation in the background of a flat class of G\"{o}del-type space-time}}

\end{center}

\vspace{.5cm}

\begin{center}
{\bf Faizuddin Ahmed}\footnote {faizuddinahmed15@gmail.com (Corresponding author)}\\ 
{\it Ajmal College of Arts and Science, Dhubri-783324, Assam, India}\\
{\bf Hassan Hassanabadi}\footnote{hha1349@gmail.com}\\
{\it Faculty of Physics, Shahrood University of Technology, P.O. Box 3619995161-316, Shahrood, Iran}
\end{center}

\vspace{.5cm}

\begin{abstract}

In this paper, we investigate the Duffin-Kemmer-Petiau (DKP) equation for spin-zero system of charge-free particles in the background of a flat class of G\"{o}del-type space-times, and evaluate the individual energy levels and corresponding wave-functions, in details.

\end{abstract}

{\it keywords:} G\"{o}del-type space-times, relativistic wave equations, energy spectrum, wave functions.

\vspace{0.1cm}

{\it PACS Number:} 04.20.Cv, 03.65.Pm, 03.65.Ge.

\section{Introduction}

The spin-$0$ and spin-$\frac{1}{2}$ particles with or without interactions have been extensively studied in relativistic quantum mechanics. However, one can use the Duffinn-Kemmer-Petiau (DKP) equation \cite{Kemmer,Duffin,Kemmer2,Petiau} which provides a good theoretical basis for spin-$ 0 $ and spin-$ 1 $ particles. The DKP equation is a direct generalization to Dirac equation of integer spin in which we replace the gamma matrices ($\gamma$) by beta matrices ($\beta^{\mu}$). This equation has been studied in atomic and condensed matter physics \cite{Gribov,Kanatchikov,Lunardi,Lunardi2,Montigny,Hulthen,Jameelt,Barnan,Richard,Kuli}. The DKP equation under different kind of potentials have been studied by various authors \cite{BB,Berk,Boztosun,Kasri,Castro,Chargui,Silva}.

The relativistic DKP equation for spin-zero particle is given by \cite{Kemmer,Duffin,Kemmer2,Petiau}
\begin{equation}
(i\,\beta^{\mu} ({\bf x})\,\partial_{\mu}-m)\,\Psi=0\quad (\hbar=1=c),
\label{1}
\end{equation}
where $\beta^{j}(j=0,1,2,3)$ are the DKP matrices which satisfy the commutation relation
\begin{equation}
\beta^{\mu}\,\beta^{\nu}\,\beta^{\lambda}+\beta^{\lambda}\,\beta^{\nu}\,\beta^{\mu}=g^{\mu\nu}\,\beta^{\lambda}+g^{\lambda\nu}\,\beta^{\mu}.
\label{2}
\end{equation}
The set of flat beta ($\beta^a$) matrices are
\begin{equation}
\beta^{0}=\left (\begin{array}{lllll} 
0 & 1 & 0 & 0 & 0 \\
1 & 0 & 0 & 0 & 0 \\
0 & 0 & 0 & 0 & 0 \\
0 & 0 & 0 & 0 & 0 \\
0 & 0 & 0 & 0 & 0
\end{array} \right),\quad
\beta^{1}=\left (\begin{array}{lllll}
0 & 0 & -1 & 0 & 0 \\
0 & 0 & 0 & 0 & 0 \\
1 & 0 & 0 & 0 & 0 \\
0 & 0 & 0 & 0 & 0 \\
0 & 0 & 0 & 0 & 0
\end{array} \right),\nonumber\\
\end{equation}
\begin{equation}
\beta^{2}=\left (\begin{array}{lllll}
0 & 0 & 0 & -1 & 0 \\
0 & 0 & 0 & 0 & 0 \\
0 & 0 & 0 & 0 & 0 \\
1 & 0 & 0 & 0 & 0 \\
0 & 0 & 0 & 0 & 0
\end{array} \right),\quad
\beta^{3}=\left (\begin{array}{lllll}
0 & 0 & 0 & 0 & -1 \\
0 & 0 & 0 & 0 & 0 \\
0 & 0 & 0 & 0 & 0 \\
0 & 0 & 0 & 0 & 0 \\
1 & 0 & 0 & 0 & 0
\end{array} \right).
\label{a1}
\end{equation}

Recently, Hassan {\it et al} studied the DKP equation in a class of spherical symmetry G\"{o}del-type space-time backgrounds \cite{Hassan}.

The general form of G\"{o}del-type metrics in polar coordinates $(t,r,\phi,z)$ can be written as \cite{Hassan,EPJC3}
\begin{equation}
ds^2=-(dt+F(r)\,d\phi)^2+H^{2}(r)\,d\phi^2+dr^2+dz^2.
\label{a2}
\end{equation}
The necessary and sufficient conditions for G\"{o}del-type solution (\ref{a2}) to be STH (space-time homogeneous) are given by
\begin{equation}
\frac{F'}{H}=2\,\Omega \quad,\quad \frac{H''}{H}=\mu^2,
\label{a3}
\end{equation}
where the prime denote derivative with respect r, and the parameters $(\Omega, \mu)$ are constants such that $\Omega^2 > 0$ and $-\infty \leq \mu^2 \leq \infty$. The parameter $\Omega$ characterizes the vorticity of the space-time. The general form of G\"{o}del-type space-time representing surfaces of constant curvature with $\mu^2 <0$, $\mu^2=0$, and $\mu^2>0$ respectively a spherical symmetry, flat and hyperbolic cases \cite{Hassan,EPJC3}. For vanishing vorticity parameter, $\Omega \rightarrow 0$, the space-time reduces to Minkowski metric in cylindrical coordinates $(t,r,\phi,z)$.

In the present work, we study the DKP equation for spin-$0$ system in a class of flat G\"{o}del-type space-time background, and evaluate the energy eigenvalues and corresponding wave function.

\section{The DKP equation in a G\"{o}del-type metric}

Consider the following stationary space-time \cite{Faiz} (see Refs. \cite{EPJC3,EPJC,EPJC2}) in the coordinates $(t,x,y,z)$ given by
\begin{eqnarray}
ds^2=-dt^2+dx^2+\left(1-\alpha_{0}^2\,x^2\right)dy^2-2\,\alpha_{0}\,x\,dt\,dy+dz^2,
\label{3}
\end{eqnarray}
where $\alpha_0>0$ is a real number.

Using the condition (\ref{a3}) into the spae-time (\ref{3}), we have $\alpha_0=2\,\Omega$ and $\mu=0$, which indicates the study space-time belong to a class of flat G\"{o}del-type metrics \cite{Hassan,EPJC3}. For $\Omega \rightarrow 0$, the study space-time reduces to Minkowski metric.

The metric tensor for the space-time (\ref{3}) be 
\begin{equation}
g_{\mu\nu} ({\bf x})=\left (\begin{array}{llll}
-1 & 0 & -\alpha_0 x & 0 \\
\quad 0 & 1 & \quad 0 & 0 \\
-\alpha_0 x & 0 & 1-\alpha_{0}^2 x^2 & 0 \\
\quad 0 & 0 & \quad 0 & 1
\end{array} \right)
\label{a4}
\end{equation}
with its inverse 
\begin{equation}
g^{\mu\nu} ({\bf x})=\left (\begin{array}{llll}
\alpha_{0}^2 x^2-1 & 0 & -\alpha_0 x & 0 \\
\quad\quad 0 & 1 & \quad 0 & 0 \\
\quad-\alpha_0 x & 0 & \quad 1 & 0 \\
\quad\quad 0 & 0 & \quad 0 & 1
\end{array} \right).
\label{a5}
\end{equation}

In order to work with spin-0 scalar particles by solving the DKP equation, let us define the local reference frame via a non-coordinate basis $\hat {\theta}^{a}=e^{a}_{\mu}({\bf x})\,dx^{\mu}$, where the components $e^{a}_{\mu} ({\bf x})$ are called tetrads and satisfy the relation \cite{SW,PB,CWM,MN}:
\begin{equation}
g_{\mu\nu} ({\bf x})= e^{a}_{\mu} ({\bf x})\,e^{b}_{\nu} ({\bf x})\,\eta_{ab},
\label{b1}
\end{equation}
where $\eta_{ab}=\mbox{diag} (-,+,+,+)$ is the Minkowski tensor. The inverse of tetrads is defined as $dx^{\mu}=e^{\mu}_{a} ({\bf x})\,\hat{\theta}^{a}$, where the following relations must satisfy:
\begin{equation}
e^{\mu}_{a} ({\bf x})\,e^{a}_{\nu} ({\bf x})=\delta^{\mu}_{\nu}\quad,\quad e^{a}_{\mu} ({\bf x})\,e^{\mu}_{b} ({\bf x})=\delta^{a}_{b}.
\label{b2}
\end{equation}
Observe that the indices $\mu,\nu$ indicate the space-time indices, while $a,b=0,1,2,3$ indicate local reference frame indices.

For the space-time (\ref{3}), the corresponding local reference frame can be written in the form:
\begin{equation}
\hat{\theta}^0=dt+\alpha_0\,x\,dy,\quad \hat{\theta}^1=dx,\quad \hat{\theta}^2=dy,\quad \hat{\theta}^3=dz
\label{b3}
\end{equation}
so that the line-element is define as
\begin{equation}
ds^2=-(\hat{\theta}^0)^2+(\hat{\theta}^1)^2+(\hat{\theta}^2)^2+(\hat{\theta}^3)^2=\eta_{ab}\,\hat{\theta}^a\,\hat{\theta}^b.
\label{b4}
\end{equation}
Thus the tetrads $e^{\mu}_{a}$ and $e^{a}_{\mu}$ for the space-time (\ref{3}) is
\begin{equation}
e^{a}_{\mu} ({\bf x})=\left (\begin{array}{llll}
1 & 0 & \alpha_0\,x & 0 \\
0 & 1 & \,0 & 0 \\
0 & 0 & \,1 & 0 \\
0 & 0 & \,0 & 1
\end{array} \right),\nonumber
\end{equation}
\begin{equation}
e^{\mu}_{a} ({\bf x})=\left (\begin{array}{llll}
1 & 0 & -\alpha_0\,x & 0 \\
0 & 1 & \quad 0 & 0 \\
0 & 0 & \quad 1 & 0 \\
0 & 0 & \quad 0 & 1
\end{array} \right).
\label{a7}
\end{equation}

The spin connections $\omega_{\mu\,ab}$, and the Christoffel symbols $\Gamma^{\sigma}_{\mu\nu} $ for the above space-time are given in Ref. \cite{EPJC3}. The non-vanishing component of spinorial affine connection $\Gamma_{\mu} (\bf x)$, according to the definition $\Gamma_{\mu} {(\bf x)}=\frac{1}{2}\,\omega_{\mu\,ab} {(\bf x)}\,[\beta^{a}, \beta^{b}]$, can be identified as
\begin{equation}
\Gamma_{t} ({\bf x})=\left (\begin{array}{lllll}
0 & 0 & 0 & \quad 0 & 0 \\
0 & 0 & 0 & \quad 0 & 0 \\
0 & 0 & 0 & -\frac{\alpha_0}{2} & 0 \\
0 & 0 & \frac{\alpha_0}{2} & \quad 0 & 0 \\
0 & 0 & 0 & \quad 0 & 0
\end{array} \right),\nonumber
\end{equation}
\begin{equation}
\Gamma_{x} ({\bf x})=\left (\begin{array}{lllll}
0 & \quad 0 & 0 & 0 & 0 \\
0 & \quad 0 & \frac{\alpha_0}{2} & 0 & 0 \\
0 & \quad 0 & 0 & 0 & 0 \\
0 & -\frac{\alpha_0}{2} & 0 & 0 & 0\\
0 & \quad 0 & 0 & 0 & 0
\end{array} \right),\nonumber
\end{equation}
\begin{equation}
\Gamma_{y} ({\bf x})=\left (\begin{array}{lllll}
0 & \quad 0 & \quad 0 & \quad 0 & 0 \\
0 & \quad 0 & \quad \frac{\alpha_0}{2} & \quad 0 & 0 \\
0 & -\frac{\alpha_0}{2} & \quad 0 & -\alpha_{0}^2\,x & 0 \\
0 & \quad 0 & -\alpha_{0}^2\,x & \quad 0 & 0 \\
0 & \quad 0 & \quad 0 & \quad 0 & 0
\end{array} \right),\nonumber
\end{equation}
\begin{equation}
\Gamma_{z} ({\bf x})=\left (\begin{array}{lllll}
0 & 0 & 0 & 0 & 0 \\
0 & 0 & 0 & 0 & 0 \\
0 & 0 & 0 & 0 & 0 \\
0 & 0 & 0 & 0 & 0 \\
0 & 0 & 0 & 0 & 0
\end{array} \right).
\label{7}
\end{equation}
The kemmer matrices $\beta^{\mu} ({\bf x})=e^{\mu}_{a} ({\bf x})\,\beta^{a}$ in the curved space-time are
\begin{equation}
\beta^{t} ({\bf x})=\left(\begin{array}{lllll} 
0 & 1 & 0 & 0 & 0 \\
1 & 0 & 0 & 0 & 0 \\
0 & 0 & 0 & 0 & 0 \\
0 & 0 & 0 & 0 & 0 \\
0 & 0 & 0 & 0 & 0
\end{array} \right),\quad
\beta^{x} ({\bf x})=\left (\begin{array}{lllll}
0 & 0 & -1 & 0 & 0 \\
0 & 0 & \,\,0 & 0 & 0 \\
1 & 0 & \,\,0 & 0 & 0 \\
0 & 0 & \,\,0 & 0 & 0 \\
0 & 0 & \,\,0 & 0 & 0
\end{array} \right),\nonumber
\end{equation}
\begin{equation}
\beta^{y} ({\bf x})=\left (\begin{array}{lllll}
\quad 0 & -\alpha_0\,x & \quad 0 & -1 & 0 \\
\alpha_0\,x & \quad 0 & \quad 0 & \quad 0 & 0 \\
\quad 0 & \quad 0 & \quad 0 & \quad 0 & 0 \\
\quad 1 & \quad 0 & \quad 0 & \quad 0 & 0 \\
\quad 0 & \quad 0 & \quad 0 & \quad 0 & 0
\end{array} \right),\,
\beta^{z} ({\bf x})=\left (\begin{array}{lllll}
0 & 0 & 0 & 0 & \,-1 \\
0 & 0 & 0 & 0 & \quad 0 \\
0 & 0 & 0 & 0 & \quad 0 \\
0 & 0 & 0 & 0 & \quad 0 \\
1 & 0 & 0 & 0 & \quad 0
\end{array} \right).
\label{9}
\end{equation}

The DKP equation in curved space is given by
\begin{equation}
[i\,\beta^{\mu} ({\bf x})\,(\partial_{\mu}+\Gamma_{\mu} ({\bf x}))-m]\,\Psi=0,
\label{10}
\end{equation}
where  $\beta^{\mu} (x)$ and $\Gamma_{\mu} (x)$ are derived above. We have
\begin{equation}
\beta^{\mu} ({\bf x})\,\Gamma_{\mu} ({\bf x})=-\frac{3\alpha_{0}^2\,x}{2}\,\left (\begin{array}{lllll}
0 & 0 & 1 & 0 & 0 \\
0 & 0 & 0 & 0 & 0 \\
0 & 0 & 0 & 0 & 0 \\
0 & 0 & 0 & 0 & 0 \\
0 & 0 & 0 & 0 & 0
\end{array} \right).
\label{11}
\end{equation}
Since, the given metric (\ref{3}) is independent of $t,y,z$. Suppose the general wave function to be
\begin{equation}
\Psi (t,x,y,z)=e^{i\,(-E\,t+l\,y+k\,z)}\,\left (\begin{array}{c}
\psi_{1} (x)\\
\psi_{2} (x)\\
\psi_{3} (x)\\
\psi_{4} (x)\\
\psi_{5} (x)
\end{array} \right),
\label{12}
\end{equation}
where $E$ is the energy eigenvalues, and $l, k$ are the constants. Substituting the above ansatz (\ref{12}) into the Eq. (\ref{10}) and using (\ref{11}), we obtain the following set of differential equation:
\begin{eqnarray}
\label{13}
&&E\,\psi_{2}-i\,\psi'_{3}+l\,(\alpha_0\,x\,\psi_{2}+\psi_{4})+k\,\psi_{5}-i\,\frac{3\,\alpha_{0}^2\,x}{2}\,\psi_{3}=m\,\psi_{1},\\
\label{14}
&&E\,\psi_{1}+\alpha_0\,l\,x\,\psi_{1}=m\,\psi_{2},\\
\label{15}
&&i\,\psi'_{1}=m\,\psi_{3},\\
\label{16}
&&-l\,\psi_{1}=m\,\psi_{4},\\
\label{17}
&&-k\,\psi_{1}=m\,\psi_{5},
\end{eqnarray}
in which a prime means ordinary derivative w. r. t. $x$. The above Eqs. (\ref{14})-(\ref{17}) can now be express as
\begin{eqnarray}
\label{18}
&&\psi_{2}=\frac{1}{m}\,(E+\alpha_0\,l\,x)\,\psi_{1},\\
\label{19}
&&\psi_{3}=\frac{i}{m}\,\psi'_{1},\\
\label{20}
&&\psi_{4}=-\frac{l}{m}\,\psi_{1},\\
\label{21}
&&\psi_{5}=-\frac{k}{m}\,\psi_{1}.
\end{eqnarray}
Substituting the above Eqs. (\ref{18})-(\ref{21}) into the Eq. (\ref{13}), one will obtain the following second order linear differential equation for $\psi_{1} (x)$ :
\begin{equation}
\psi''_{1}+a\,x\,\psi'_{1}+b\,x\,\psi_{1}+c\,x^2\,\psi_{1}=\lambda\,\psi_{1},
\label{22}
\end{equation}
where 
\begin{eqnarray}
&&a=\frac{3\alpha_{0}^2}{2}=6\,\Omega^2,\nonumber\\
&&b=2\,\alpha_0\,E\,l=4\,\Omega\,l\,E,\nonumber\\
&&c=\alpha_{0}^2\,l^{2}=4\,\Omega^2\,l^2,\nonumber\\
&&\lambda=m^2+l^2+k^2-E^2.
\label{23}
\end{eqnarray}

Let us substitute
\begin{equation}
\psi_{1} (x)=e^{-\frac{a}{4}\,x^2}\,u (x),
\label{24}
\end{equation}
into the Eq. (\ref{22}), we get the following differential equation 
\begin{equation}
u''-(A\,x^2-B\,x)\,u-(\lambda+\frac{a}{2})\,u=0,
\label{25}
\end{equation}
where 
\begin{equation}
A=\frac{a^2}{4}-c\quad,\quad B=b.
\label{26}
\end{equation}
The above Eq. (\ref{25}) can be express as
\begin{equation}
u''-(\sqrt{A}\,x-\frac{B}{2\,\sqrt{A}})^2\,u+[\frac{B^2}{4\,A}-(\lambda+\frac{a}{2})]\,u=0
\label{27}
\end{equation}
Replacing 
\begin{equation}
s=\sqrt{A}\,x-\frac{B}{2\,\sqrt{A}}\quad,\quad D=\frac{B^2}{4\,A}-(\lambda+\frac{a}{2}),
\label{28}
\end{equation}
into the Eq. (\ref{27}), we obtain
\begin{equation}
u'' (s)-\frac{s^2}{A}\,u (s)+\frac{D}{A}\,u (s)=0.
\label{29}
\end{equation}
Again replacing $s$ into new variable $r$ by 
\begin{equation}
s=A^{\frac{1}{4}}\,r
\label{30}
\end{equation}
into the above Eq. (\ref{29}), we get
\begin{equation}
\frac{d^2 u}{dr^2}+(\epsilon-r^2)\, u =0,
\label{31}
\end{equation}
where $\epsilon=\frac{D}{\sqrt{A}}$. The Eq. (\ref{31}) is similar with the harmonic oscillator equation, therefore, the energy eigenvalues equation :
\begin{eqnarray}
&&\epsilon=(2\,n+1)\Rightarrow \frac{D}{\sqrt{A}}=(2\,n+1)\nonumber\\
&&\Rightarrow\frac{\frac{B^2}{4\,A}-(\lambda+\frac{a}{2})}{\sqrt{\frac{a^2}{4}-c}}=(2\,n+1)\nonumber\\
&&\Rightarrow\frac{\frac{b^2}{4\,(\frac{a^2}{4}-c)}-(\lambda+\frac{a}{2})}{\sqrt{\frac{a^2}{4}-c}}=(2\,n+1),
\label{32}
\end{eqnarray}
where $n=0,1,2,3,4,........$. Substituting the various quantities, we get the energy eigenvalues $E_{n,l}$ associated with $n^{th}$ radial modes:
\begin{equation}
    E_{n,l}=\pm\eta\,[3\,\Omega^2\,\{1+\eta\,(2\,n+1)\}+m^2+l^{2}+k^{2}]^{\frac{1}{2}},
    \label{33}
\end{equation}
where we define
\begin{equation}
    \eta=\sqrt{1-\frac{4l^{2}}{9\Omega^2}}.
    \label{34}
\end{equation}

The corresponding wave functions are given by
\begin{equation}
\psi_{1\,n} (x)=|N| e^{-\frac{3\,\Omega^2\,x^2}{2}}\,H_{n} (x),
\label{35}
\end{equation}
where $|N|=\sqrt{\frac{\sqrt{3}\,\Omega}{\sqrt{\pi}\,2^{n}\,{n!}}}$ is the normalization constant and $H_{n} (x)$ are the Hermite polynomials and define as
\begin{equation}
H_{n} (x)=(-1)^{n}\,e^{x^2}\,\frac{d^{n}}{d{x^n}}\,e^{-x^2},\quad \int^{\infty}_{-\infty} e^{-x^2}\,H_{n} (x)\,H_{m} (x)\,dx=\sqrt{\pi}\,2^{n}\,{n!}\,\delta_{nm}.
\label{36}
\end{equation}
Therefore from Eqs. (\ref{18})--(\ref{21}), we have
\begin{eqnarray}
\label{37}
&&\psi_{2\,n}=\frac{1}{m}\,(E_n+2\Omega lx)\,\psi_{1\,n},\\
\label{38}
&&\psi_{3\,n}=\frac{i}{m}\,\psi'_{1\,n},\\
\label{39}
&&\psi_{4\,n}=-\frac{l}{m}\,\psi_{1\,n},\\
\label{40}
&&\psi_{5\,n}=-\frac{k}{m}\,\psi_{1\,n}.
\end{eqnarray}
Therefore, we have the general wave function
\begin{equation}
\Psi_{n} (t,x,y,z)=\frac{1}{m}\,e^{i\,(-E_{n,l}\,t+l\,y+k\,z)}\,\left (\begin{array}{c}
m\,\psi_{1\,n} (x)\,\\
E_{n,l}\,\psi_{1\,n} (x)\,\\
i\,\psi'_{1\,n} (x)\,\\
-l\,\psi_{1\,n}\\
-k\,\psi_{1\,n}
\end{array} \right).
\label{41}
\end{equation}

It is worthwhile mentioning in Ref. \cite{EPJC} that we have solved the Klein-Gordon equation on the same space-time (\ref{3}), and evaluated the energy eigenvalues (setting $k=0=l$) as follow:
\begin{equation}
    E_{n,0}=(2\,n+1)\,\Omega+\sqrt{(2\,n+1)^2\,\Omega^2+m^2},\quad n=1,2,....
    \label{43}
\end{equation}
For $k=0=l$ from Eq (\ref{33}), we have the following energy eigenvalues
\begin{equation}
    E_{n,0}=\sqrt{6\,\Omega^2\,(n+1)+m^2},\quad n=0,1,2,.....
    \label{44}
\end{equation}
It is clear from above that the energy eigenvalues (\ref{43}) of a relativistic quantum scalar particle by solving the Klein-Gordon equation on the space-time (\ref{3}) in Ref. \cite{EPJC} are different from the eigenvalues (\ref{44}) obtained here. Another important difference is that for the Klein-Gordon equation, the ground state energy eigenvalues are obtained by setting $n=1$, and the ground state energy levels are $E_1=3\,\Omega+\sqrt{9\,\Omega^2+m^2}$. Whereas for the DKP equations, the ground state energy eigenvalues are obtained by setting $n=0$, and the ground state energy levels are $E_0=\sqrt{6\,\Omega^2+m^2}$.

Let us study the eigenvalues Eq. (\ref{33}) and corresponding wave functions Eq. (\ref{41}) one by one. We have set the constant $k = 0$.

\begin{eqnarray}
(i)\quad n=0: &&E_{0,l}=\pm \eta\sqrt{3\Omega^2 (1+ \eta)+l^2+m^2},\nonumber\\ &&\psi_{1\,0}=(\frac{3\,\Omega^2}{\pi})^{\frac{1}{4}}\,e^{-\frac{3\,\Omega^2\,x^2}{2}},\nonumber\\
&&\Psi_{0} (t,x,y)=\frac{1}{m}\,e^{-i\,E_{0}\,t+ i\,l\,y}\,\psi_{1\,0}\,\left (\begin{array}{c}
m\\
E_{0}\\
-3\,i\,\Omega^2\,x\\
-l\\
0
\end{array} \right).\nonumber\\
(ii)\quad n=1: && E_{1,l}=\pm \eta\sqrt{3\Omega^2 (1+3\eta )+l^2+m^2},\nonumber\\ &&\psi_{1\,1}=2\,(\frac{3\,\Omega^2}{4\,\pi})^{\frac{1}{4}}\,x\,e^{-\frac{3\,\Omega^2\,x^2}{2}},\nonumber\\
&&\Psi_{1} (t,x,y)=\frac{\sqrt{2}}{m}\,e^{-i\,E_{1}\,t+ i\,l\,y}\,\psi_{1\,0}\,\left (\begin{array}{c}
m\,x\\
E_{1}\,x\\
i\,(1-3\,\Omega^2\,x^2)\\
-l\\
0
\end{array} \right).\nonumber
\end{eqnarray}
\begin{eqnarray}
(iii)\quad n=2: &&E_{2,l}=\pm \eta\sqrt{3\Omega^2 (1+ 5\eta )+l^2+m^2},\nonumber\\ &&\psi_{1\,2}=(\frac{3\,\Omega^2}{4\,\pi})^{\frac{1}{4}}\,(2\,x^2-1)\,e^{-\frac{3\,\Omega^2\,x^2}{2}},\nonumber\\
&&\Psi_{2} (t,x,y)=\frac{e^{-i\,E_{2}\,t+ i\,l\,y}\,\psi_{1\,0}}{\sqrt{2}\,m}\,\left (\begin{array}{c}
m\,(2\,x^2-1)\\
E_{2}\,(2\,x^2-1)\\
i\,((4+3\,\Omega^2)\,x-6\,x^3\,\Omega^2)\\
-l\\
0
\end{array} \right).\quad\quad
\label{45}
\end{eqnarray}

If one includes a non-minimal coupling ($\xi$) of the gravitational field with the background curvature, then the DKP equation in curved space is given by
\begin{equation}
[i\,\beta^{\mu} ({\bf x})\,(\partial_{\mu}+\Gamma_{\mu} ({\bf x}))-m-\xi\,R]\,\Psi=0,
\label{49}
\end{equation}
where $R$ is the Ricci scalar and $\xi$ is a coupling constant.

For the considered metric (\ref{3}), we get the the following equations using (\ref{49}) as:
\begin{eqnarray}
\label{50}
&&\psi_{2}=\frac{1}{\tilde{m}}\,(E+\alpha_0 l x)\,\psi_{1},\\
\label{51}
&&\psi_{3}=\frac{i}{\tilde{m}}\,\psi'_{1},\\
\label{52}
&&\psi_{4}=-\frac{l}{\tilde{m}}\,\psi_{1},\\
\label{53}
&&\psi_{5}=-\frac{k}{\tilde{m}}\,\psi_{1},\\
\label{54}
&&\psi''_{1}+a\,x\,\psi'_{1}+b\,x\,\psi_{1}+c\,x^2\,\psi_{1}=\tilde{\lambda}\,\psi_{1},
\end{eqnarray}
where $a,b,c$ are given earlier and
\begin{equation}
\tilde{\lambda}={\tilde m}^2+l^{2}+k^{2}-E^2,\quad \tilde{m}=m+\xi\,R,\quad R=2\,\Omega^2.
\label{55}
\end{equation}

Following the same technique as done above, one can convert Eq. (\ref{54}) into a harmonic oscillator equation.  The energy eigenvalues are
\begin{equation}
E_{n,l}=\pm\,\eta\,[3\,\Omega^2\,\{1+\eta\,(2\,n+1)\}+(m+2\,\xi\,\Omega^2)^2+l^{2}+k^{2}]^{\frac{1}{2}}.
\label{56}
\end{equation}
From Eqs. (\ref{50})--(\ref{53}) we have
\begin{eqnarray}
\label{57}
&&\psi_{2\,n}=\frac{E_n}{\tilde{m}}\,\psi_{1\,n},\\
\label{58}
&&\psi_{3\,n}=\frac{i}{\tilde{m}}\,\psi'_{1\,n},\\
\label{59}
&&\psi_{4\,n}=-\frac{l}{\tilde{m}}\,\psi_{1\,n},\\
\label{60}
&&\psi_{5\,n}=-\frac{k}{\tilde{m}}\,\psi_{1\,n}.
\end{eqnarray}
And the general wave function
\begin{equation}
\Psi_{n} (t,x,y,z)=\frac{1}{\tilde{m}}\,e^{i (-E_{n}\,t+l\,y+k\,z)}\,\left (\begin{array}{c}
\tilde{m}\,\psi_{1\,n} (x)\,\\
E_{n}\,\psi_{1\,n} (x)\,\\
i\,\psi'_{1\,n} (x)\,\\
-l \psi_{1\,n} (x)\\
-k \psi_{1\,n} (x)
\end{array} \right).
\label{61}
\end{equation}
where $\psi_{1\,n} (x)$ is given by Eq. (\ref{41}). 

The eigenvalues and corresponding wave functions for $n=0,1,2$ are given below (setting $k=0$).
\begin{eqnarray}
(i)\quad n=0: &&E_{0,l}=\pm \eta \sqrt{3\Omega^2(1+\eta)+(m+2\,\xi\,\Omega^2)^2+l^2},\nonumber\\ &&\psi_{1\,0}=(\frac{3\,\Omega^2}{\pi})^{\frac{1}{4}}\,e^{-\frac{3\,\Omega^2\,x^2}{2}},\nonumber\\
&&\Psi_{0} (t,x,y)=\frac{e^{-i\,E_{0}\,t+i\,l\,y}\,\psi_{1\,0}}{(m+2\,\xi\,\Omega^2)}\,\left (\begin{array}{c}
(m+2\,\xi\,\Omega^2)\\
E_{0}\\
-3\,i\,\Omega^2\,x\\
-l\\
0
\end{array} \right).\nonumber
\end{eqnarray}
\begin{eqnarray}
(ii)\quad n=1: &&E_{1,l}=\pm \eta\sqrt{3\Omega^2(1+3\eta)+(m+2\,\xi\,\Omega^2)^2+l^2},\nonumber\\ &&\psi_{1\,1}=2\,(\frac{3\,\Omega^2}{4\,\pi})^{\frac{1}{4}}\,x\,e^{-\frac{3\,\Omega^2\,x^2}{2}},\nonumber\\
&&\Psi_{1} (t,x,y)=\frac{\sqrt{2}\,e^{-i\,E_{1}\,t+i\,l\,y}\,\psi_{1\,0}}{(m+2\,\xi\,\Omega^2)}\,\left (\begin{array}{c}
(m+2\,\xi\,\Omega^2)\,x\\
E_{1}\,x\\
i\,(1-3\,\Omega^2\,x^2)\\
-l\\
0
\end{array} \right).\nonumber\\
(iii)\quad n=2: &&E_{2,l}=\pm \eta \sqrt{3\Omega^2(1+5\eta)+(m+2\,\xi\,\Omega^2)^2+l^2},\nonumber\\ &&\psi_{1\,2}=(\frac{3\,\Omega^2}{4\,\pi})^{\frac{1}{4}}\,(2\,x^2-1)\,e^{-\frac{3\,\Omega^2\,x^2}{2}},\nonumber\\
&&\Psi_{2} (t,x,y)=\frac{e^{-i\,E_{2}\,t+i\,l\,y}\,\psi_{1\,0}}{\sqrt{2}\,(m+2\,\xi\,\Omega^2)}\,\left (\begin{array}{c}
(m+2\,\xi\,\Omega^2)\,(2\,x^2-1)\\
E_{2}\,(2\,x^2-1)\\
i\,((4+3\,\Omega^2)\,x-6\,x^3\,\Omega^2)\\
-l\\
0
\end{array} \right).\quad\quad
\label{62}
\end{eqnarray}

\section{Conclusions}

In Ref. \cite{Hassan}, the Duffin-Kemmer-Petiau (DKP) equation for spin-$0$ system in the presence of G\"{o}del-type background space-time, were studied. They derived the final form of this equation in G\"{o}del-type space-time with a spherical symmetry ($\mu^2<0$) in the presence cosmic string, evaluated the energy eigenvalues and corresponding eigenfunctions. In this work, we study the DKP equation for spin-$0$ system in G\"{o}del-type background space-time, an example of a flat cases ($\mu^2=0$) of G\"{o}del-type space-time. We have derived the final form of this equation and finally shown similar to harmonic oscillator equation after suitable transformations. We then have obtained the energy eigenvalues Eq. (\ref{33}), and corresponding wavefunctions Eq. (\ref{41}). We have seen the eigenvalues get modifies and depend on the vorticity parameter ($\Omega$) characterising the space-time. We have obtained the individual energy spectrum and corresponding wavefunctions one by one for $n=0,1,2$. In comparison to the energy eigenvalues as obtained in Ref. \cite{EPJC} for spin-$0$ particles by solving the Klein-Gordon equation on the same space-time, the energy eigenvalues Eq. (\ref{33}) are found different. Furthermore, we have introduced a non-minimal coupling ($\xi$) of the gravitational field with the background curvature $R$ (the Ricci scalar) on the DKP equation, and derived the final form of this equation. Using similar technique as done above, we have solved it and obtained the energy eigenvalues Eq. (\ref{56}) and corresponding eigenfunctions Eq. (\ref{61}). We have seen that the presence of vorticity parameter ($\Omega$), and the coupling constant ($\xi$) modifies the energy eigenvalues Eq. (\ref{56}). Here also, we obtained the individual energy levels and corresponding wavefunction one by one by for $n=0,1,2$ and others are in the same way.

In the context of quantum chromodynamics (QCD), Cosmology, and gravity and in many areas of physics, including those in particle and nuclear physics \cite{BCC,GG,RC}, the DKP equation has been examined. Besides the importance of presenting an exactly solvable model in relativistic quantum mechanics in curved space-time, the present work may find some interesting physical applications and the results may be used to study quantum field theory in curved rotating space-time.

\section*{Acknowledgements}
We sincerely acknowledge valuable comments and suggestions from the anonymous kind referee(s).

\end{document}